\documentclass{ws-ijmpa}

\begin{document}

%

\def\nocropmarks{\vskip5pt\phantom{cropmarks}}

\let\trimmarks\nocropmarks      

%

\markboth{Edmond L. Berger}
{The Puzzle of the Bottom Quark Production Cross Section}

%
\catchline{}{}{}

%

\setcounter{page}{1}

\title{THE PUZZLE OF THE BOTTOM QUARK\\ PRODUCTION CROSS SECTION}

\author{\footnotesize EDMOND L. BERGER}

\address{High Energy Physics Division, Argonne National Laboratory\\
Argonne, Illinois 60439, USA}

\maketitle
  \vspace*{-7.2cm}
  \begin{flushright}
  \noindent hep-ph/0201229 \\
  ANL-HEP-PR-02-001 \\
  \end{flushright}
  \vspace*{+6.0cm}

\begin{abstract}
The production rate of bottom quarks at hadron colliders exceeds the 
expectations of next-to-leading order perturbative quantum chromodynamics.  
An additional contribution from pair-production of light gluinos, of mass 
12 to 16 GeV, with two-body decays into bottom quarks and light bottom 
squarks, yields a differential cross section for bottom quarks in 
better agreement with data.  The masses of the gluino and bottom squark are 
restricted further by the ratio of like-sign to opposite-sign leptons at 
hadron colliders.  Restrictions on this scenario from other data are summarized, 
and predictions are made for other processes such as Upsilon decay into a 
pair of bottom squarks.  
\end{abstract}

\section{Introduction}
The cross section for bottom-quark production at hadron collider energies exceeds 
the central value of predictions of next-to-leading order (NLO) perturbative quantum 
chromodynamics (QCD) by about a factor of
two.\cite{expxsec,ua1data}  This longstanding discrepancy has resisted fully 
satisfactory resolution within the standard model.\cite{qcdrev}  The NLO 
contributions are large, and it is not excluded that a combination of further 
higher-order effects in production and/or fragmentation 
may resolve the discrepancy.  However, the disagreement is surprising because 
the relatively large mass of the bottom quark sets a hard scattering scale at
which fixed-order perturbative QCD computations of other processes are generally
successful.  The photoproduction cross section at DESY's HERA\cite{herab} and 
and the cross section in photon-photon reactions at CERN's LEP\cite{lepb} also 
exceed NLO expectations.  The data invite the possibility of a 
contribution from ``new physics".\cite{ourletter}  

\section{Supersymmetry Interpretation}
The properties of new particles that can contribute significantly to the bottom 
quark ($b$) cross section are fairly well circumscribed.  To be produced with enough 
cross section the particles must interact strongly and have relatively low mass.  
They must either decay into $b$ quarks or be close imitators of $b$'s in a 
variety of channels of observation.  They must evade constraints 
based on precise data from measurements of $Z^o$ decays at CERN's LEP and SLAC's 
SLC, and from many lower-energy $e^+ e^-$ collider experiments.  The minimal 
supersymmetric standard model (MSSM) is a favorite candidate for 
physics beyond the standard model.  It offers a well-motivated theoretical 
framework and is reasonably well-explored phenomenologically.  

An explanation within the context of the MSSM can satisfy all of the stated 
criteria.\cite{ourletter}  
The existence is assumed of a relatively light color-octet gluino 
$\tilde g$ (mass $\simeq 12$ to 16 GeV) that decays with 100\% branching 
fraction into a bottom quark $b$ and a light color-triplet bottom squark 
$\tilde b$ (mass $\simeq 2$ to 5.5 GeV).  The $\tilde g$ and the $\tilde b$ are 
the spin-1/2 and spin-0 supersymmetric partners of the gluon ($g$) and bottom 
quark.  In this scenario the $\tilde b$ is the lightest SUSY particle, and the 
masses of all other SUSY particles are arbitrarily heavy, i.e., of order the 
electroweak scale or greater.  (The $\tilde b$ either lives long enough to 
escape from a typical collider detector or decays promptly via 
R-parity violation into a pair of hadronic jets.)  Improved agreement is obtained 
with hadron collider rates of bottom-quark production, and several predictions are 
made that can be tested readily with forthcoming data. 
 
\subsection{Differential Cross Section}
The light gluinos are produced in pairs via standard QCD subprocesses, dominantly 
$g + g \rightarrow \tilde g + \tilde g$ at Tevatron and Large Hadron Collider (LHC) 
energies.  The $\tilde g$ has
a strong color coupling to $b$'s and $\tilde b$'s and, as long as its mass
satisfies $m_{\tilde g} > m_b + m_{\tilde b}$, the $\tilde g$ decays
promptly to $b + \tilde b$.  The magnitude of the $b$ cross section, the
shape of the $b$'s transverse momentum $p_{Tb}$ distribution, and the CDF
measurement\cite{cdfmix} of $B^0 - \bar B^0$ mixing are three features of
the data that help to establish the preferred masses of the $\tilde g$ and
$\tilde b$.  Shown in 
Fig.~1 is the integrated $p_{Tb}$ distribution of the $b$ quarks that results
from $\tilde g \rightarrow b + \tilde b$, for $m_{\tilde g} = $14 GeV and
$m_{\tilde b} =$ 3.5 GeV.  The results are compared with the cross section
obtained from next-to-leading order (NLO) perturbative QCD
and CTEQ4M parton distribution functions (PDF's),\cite{cteq} with $m_b =$
4.75 GeV, and a renormalization and factorization scale $\mu=\sqrt{m_b^2 +
p_{Tb}^2}$.  SUSY-QCD corrections to $b \bar{b}$ production are not
included as they are not available and are generally expected to be 
smaller than the standard QCD corrections.  The $\tilde g$-pair cross 
section is computed from the
leading order (LO) matrix element with NLO PDF's,\cite{cteq}
$\mu=\sqrt{m^2_{\tilde{g}} + p^2_{T_{\tilde{g}}}}$, and a two-loop 
expression for  the strong coupling $\alpha_s$. To account for NLO effects, 
this $\tilde g$-pair cross section is multiplied by 1.9, the ratio 
of inclusive NLO to LO cross sections.\cite{prospino}

A relatively light gluino is necessary in order to obtain a bottom-quark
cross section comparable in magnitude to the pure QCD component.  Values of
$m_{\tilde g} \simeq$ 12 to 16 GeV are chosen because the resulting $\tilde
g$ decays produce $p_{Tb}$ spectra that are enhanced primarily in the
neighborhood of $p_{Tb}^{\rm min} \simeq m_{\tilde g}$ where the data show
the most prominent enhancement above the QCD expectation.  Larger values of
$m_{\tilde g}$ yield too little cross section to be of interest, and
smaller values produce more cross section than seems tolerated by the ratio
of like-sign to opposite-sign leptons from $b$ decay, as discussed below.
The choice of $m_{\tilde b}$ has an impact on the kinematics of the $b$.
After selections on $p_{Tb}^{\rm min}$, large values of $m_{\tilde b}$
reduce the cross section and, in addition, lead to shapes of the $p_{Tb}$
distribution that agree less well with the data.  

\begin{figure}[htbp]
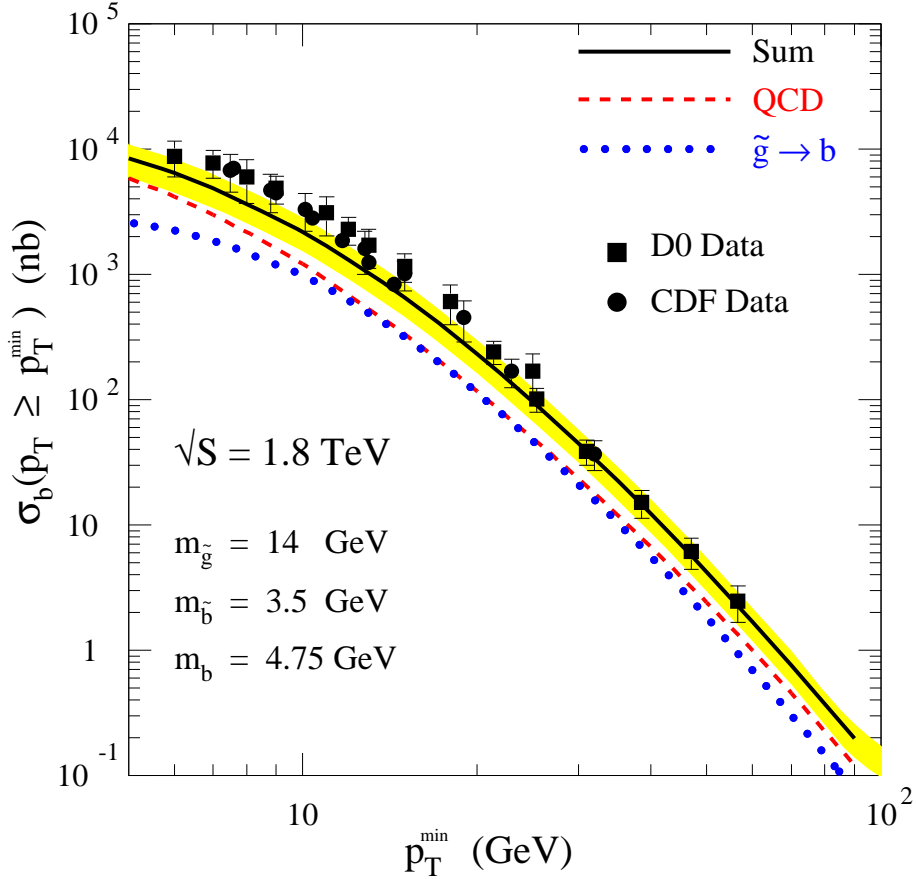
 
\figurebox{}{}{berf1} 
\caption{Bottom-quark cross section in $p\bar p$ collisions at $\sqrt{S}
=1.8$ TeV for $p_{Tb}>p_{Tb}^{\rm min}$ with a gluino of mass
$m_{\tilde{g}} = 14$ GeV and a bottom squark of mass $m_{\tilde{b}} = 3.5$
GeV.  The dashed curve is the central value of the NLO QCD prediction. The
dotted curve shows the $p_T$ spectrum of the $b$ from the SUSY processes.  The 
solid curve is the sum of the QCD and SUSY components.  The shaded band 
represents an uncertainty of roughly $\pm$30\% associated with variations of 
the renormalization and factorization scales, the $b$ mass, and the parton 
densities.  The rapidity cut on the $b$'s is $|y_b| \le 1$. Data are from Ref.~1.}
\end{figure}

After the contributions of the NLO QCD and SUSY components are added, 
the magnitude of the bottom-quark cross section and the shape of the integrated 
$p^{\rm min}_{Tb}$ distribution are described well. Very good agreement is 
obtained also with data from the UA1 experiment (not shown).\cite{ua1data}
The SUSY process produces bottom quarks in a four-body final state and thus
their momentum correlations are different from those of QCD.  Angular
correlations between muons that arise from decays of $b$'s have been
measured.\cite{cdfmix,muonexp}  The angular correlations between
$b$'s in the SUSY case are nearly indistinguishable from those
of QCD once experimental cuts are applied.    

The energy dependence of the bottom cross section is a potentially
important constraint on models in which new physics is invoked to
interpret the observed excess bottom quark yield.  Since the 
assumed $\tilde{g}$ mass is larger than the mass
of the $b$, the $\tilde{g}$ pair process will turn on more slowly with
energy than pure QCD production of $b \bar{b}$ pairs.  The new physics
contribution will depress the ratio of cross sections at 630 GeV and
1.8 TeV from the pure QCD expectation.  An explicit calculation with 
CTEQ4M parton densities and the $b$ rapidity selection $|y| < 1$, yields 
a pure QCD prediction at NLO of 0.17 +/- 0.02 for 
$p_{Tb}^{\rm min} =$ 10.5 GeV, and 0.16 +/- 0.02 
after inclusion of the gluino pair contribution.  Either of these numbers 
is consistent with forthcoming data from CDF on this ratio.\cite{privcom} 

\subsection{Same-sign to Opposite-sign Leptons}
Since the $\tilde g$ is a Majorana particle, its decay yields both quarks
and antiquarks.  Gluino pair production and subsequent decay to $b$'s will
generate $b b$ and $\bar b \bar b$ pairs, as well as the $b \bar b$
final states that appear in QCD production.  When a gluino is highly 
relativistic, its helicity is nearly the same as its chirality.  Therefore, 
selection of $\tilde g$'s whose transverse momentum is greater than their mass 
will reduce the number of like-sign $b$'s.  In the intermediate $p_T$ region,
however, the like-sign suppression is reduced.  The cuts chosen in current 
hadron collider experiments for measurement of the ratio of like-sign to 
opposite-sign muons result in primarily unpolarized $\tilde g$'s, and, 
independent of the
$\tilde b$ mixing angle, an equal number of like-sign and opposite-sign
$b$'s is expected at production.  The SUSY mechanism leads therefore to an 
increase of like-sign leptons in the final state after semi-leptonic decays of 
the $b$ and $\bar b$ quarks.  This increase could be confused with an enhanced 
rate of $B^0-\bar B^0$ mixing.  

Time-integrated mixing analyses of lepton pairs 
observed at hadron colliders are interpreted in terms of the 
quantity $\bar{\chi} = f_d \chi_d + f_s \chi_s$, where 
$f_d$ and $f_s$ are the fractions of
$B^0_d$ and $B^0_s$ hadrons in the sample of semi-leptonic $B$
decays, and $\chi_f$ is the time-integrated mixing probability for $B^0_f$.
Conventional $b\bar b$ pair production determines the quantity 
$LS_c = 2\bar{\chi} (1-\bar{\chi})$, the fraction of $b\bar b$ pairs that 
decay into like-sign leptons.  The SUSY mechanism leads to a new 
expression  
\begin{equation}
LS =\frac{1}{2} \frac{\sigma_{\tilde{g}\tilde{g}}}{\sigma_{\tilde{g}\tilde{g}}+
\sigma_{\rm{qcd}}} + 
LS_c \frac{\sigma_{\rm {qcd}}}{\sigma_{\tilde{g}\tilde{g}}+\sigma_{\rm {qcd}}} = 
2 \bar{\chi}_{\rm {eff}} (1 - \bar{\chi}_{\rm {eff}}).  
\end{equation}
The factor $1/2$ arises because $N(bb + \bar{b}\bar{b}) \simeq N(b \bar{b})$ in 
the SUSY mechanism for the selections on $p_{Tb}$ made in the CDF run I analysis.  
Defining $G = \sigma_{\tilde{g}\tilde{g}} / \sigma_{\rm{qcd}}$, 
the ratio of SUSY and QCD bottom-quark cross sections after cuts, and solving for 
the effective mixing parameter, one obtains
\begin{equation}
\bar{\chi}_{\rm {eff}}=\frac{\bar{\chi}}{\sqrt{1+G}} +{1\over 2}\left[1-
	\frac{1}{\sqrt{1+G}}\right] .
\end{equation}
The CDF measurement\cite{cdfmix} of 
$\bar{\chi}_{\rm {eff}} = 0.131 \pm 0.02 \pm 0.016$   
is marginally larger than the world average value $\bar{\chi} = 
0.118 \pm 0.005$,\cite{pdg} assumed to be the contribution from the pure QCD 
component only.  

The ratio $G$ is determined in the region of phase space where the measurement is 
made,\cite{cdfmix} with both final $b$'s having $p_{Tb} \ge 6.5$ GeV and rapidity 
$| y_b | \leq 1$.  With $m_{\tilde b} =$ 3.5 GeV, $G =$ 0.37 and 0.28 for gluino 
masses $m_{\tilde g} =$ 14 and 16 GeV, respectively.  The predictions are 
$\bar{\chi}_{\rm {eff}} = 0.17 \pm 0.02 $ for $m_{\tilde g} =$ 14 GeV, and 
$\bar{\chi}_{\rm {eff}} = 0.16 \pm 0.02 $ with $m_{\tilde g} =$ 16 GeV.
Additional theoretical uncertainties
arise because there is no fully differential NLO calculation of gluino
production and subsequent decay to $b$'s.  The choice $m_{\tilde g} > 12$ GeV 
leads to a calculated $\bar{\chi}_{\rm {eff}}$ consistent with the data within 
uncertainties.  With $\sigma_{\tilde{g}\tilde{g}} / \sigma_{\rm{qcd}} \sim 1/3$, 
the mixing data and the magnitude and $p_T$ dependence of the $b$ production 
cross section can be satisfied.  

\section{Other Experimental and Theoretical Constraints}
An early study by the UA1 Collaboration\cite{ua1gluino} 
excludes $\tilde{g}$'s in the mass range $4 < m_{\tilde{g}} 
< 53$ GeV, but it starts from the assumption that there is a light neutralino 
${\tilde{\chi}}_1^0$ whose mass is less than the mass of the gluino.  The 
conclusion is based on the absence of the expected decay $\tilde{g} \rightarrow 
q+\bar{q}+\not{E_T}$, where $\not{E_T}$ represents the missing energy 
associated with the ${\tilde{\chi}}_1^0$.  In the scenario discussed above, this 
decay process does not occur since the bottom squark is the LSP, the SUSY particle 
with lowest mass, and the ${\tilde{\chi}}_1^0$ mass is presumed to be large 
({\em i.e.}, $> 50$ GeV). An analysis of 2- and 4-jet events by the ALEPH 
collaboration\cite{ALEPH} disfavors $\tilde g$'s with mass 
$m_{\tilde g} < 6.3$ GeV but not $\tilde g$'s in the mass range relevant for the 
SUSY interpretation of the bottom quark production cross section.  A similar 
analysis is reported by the OPAL collaboration.\cite{OPALg}  A light 
$\tilde{b}$ is not excluded by the ALEPH analysis.  
The exclusion by the CLEO collaboration\cite{CLEO} of a $\tilde b$ with mass 
3.5 to 4.5 GeV does not apply since their analysis focuses only on the decays 
$\tilde b \rightarrow c \ell \tilde \nu$ and 
$\tilde b \rightarrow c \ell $.  The $\tilde b$ need not decay leptonically nor 
into charm.  On the other hand, these data might be 
reinterpreted in terms of a bound on the R-parity violating lepton-number 
violating decay of $\tilde b$ into $c \ell$.  It would be interesting  
to study the hadronic decays $\tilde b \rightarrow c q$, with $q = d$ or $s$, 
and $\tilde b \rightarrow u s$ with the CLEO data.  The DELPHI 
collaboration's\cite{DELPHI} search for long-lived squarks in their 
$\gamma \gamma$ event sample is not sensitive to $m_{\tilde b} < 15$ GeV.  
The combined ranges of $\tilde{b}$ and $\tilde{g}$ masses are also compatible 
with renormalization group equation constraints and the absence of color 
and charge breaking minima in the scalar potential.\cite{DedesDreiner}

There are important restrictions on couplings of the bottom squarks from precise 
measurements of $Z^0$ decays.  A light $\tilde b$ would be ruled out unless its 
coupling to the $Z^0$ is very small.  The squark couplings to the $Z^0$ depend 
on the mixing angle $\theta_b$.  If the light 
bottom squark ($\widetilde{b}_1$) is an appropriate mixture of left-handed and 
right-handed bottom squarks, its lowest-order (tree-level) coupling to the $Z^0$ 
can be arranged to be small\cite{CHWW} if $\sin^2 \theta_b \sim 1/6$.  The couplings 
$Z_{\tilde{b}_1 \tilde{b}_2}$ and $Z_{\tilde{b}_2 \tilde{b}_2}$ survive, where 
$\widetilde{b}_2$ is the heavier bottom squark.  However, as long as the 
combination of the masses $m_{\tilde{b}_1} + m_{\tilde{b}_2}$ is less than the 
maximum center-of-mass energy explored at LEP, these couplings present no 
difficulty.  This condition implies roughly $m_{\tilde{b}_2} > 200$ GeV.  However, 
much lower masses of $\tilde{b}_2$ might be tolerated.  A careful phenomenological 
analysis is needed of expected $\widetilde{b}_2$ decay signatures, along with 
an understanding of detection efficiencies and expected event rates, before one 
knows the admissible range of masses consistent with LEP data.  At higher-order, 
unless the $\widetilde{b}_2$ mass is of order 100 GeV, contributions from loop 
processes in which light gluinos are exchanged may produce significant deviations 
from measurements of the ratios $A_b$, the forward-backward $b$ asymmetry at the 
$Z^0$, and $R_b$, the hadronic branching ratio of the $Z^0$ into $b$ 
quarks.\cite{loops1,loops2}    

Bottom squarks make a small contribution to the
inclusive cross section for $e^+ e^- \rightarrow$ hadrons, in comparison to
the contributions from quark production, and $\tilde{b} {\tilde{b}}^*$
resonances are likely to be impossible to extract from 
backgrounds.\cite{Nappi}  The angular distribution of hadronic jets produced 
in $e^+ e^-$ 
annihilation can be examined in order to bound the contribution of
scalar-quark production.  Spin-1/2 quarks and spin-0 squarks emerge with
different distributions, $(1 \pm {\rm cos}^2 \theta)$, respectively. 
The angular distribution measured by the CELLO
collaboration\cite{CELLO} is consistent with the production
of a single pair of charge-1/3 squarks along with five flavors of
quark-antiquark pairs.  Greater statistics would be valuable.

\section{Predictions and Implications}

\subsection{$\Upsilon$ Decay into Bottom Squarks}

If the bottom squark mass is less than half the mass of one of the Upsilon states,  
then Upsilon decay to a pair of bottom squarks might proceed with sufficient rate 
for experimental observation or exclusion of a light bottom squark.  The expected 
rate for $\Upsilon \rightarrow \tilde b {\tilde b}^*$ may be computed as a function 
of the masses of the bottom squark and the gluino.\cite{elblc}  

The data sample is largest at the $\Upsilon(4S)$.  For a fixed 
gluino mass of 14 GeV, the branching fraction into a pair of bottom 
squarks is about $10^{-3}$, for $m_{\tilde b} =$ 2.5 GeV, and about $10^{-4}$ for 
$m_{\tilde b} =$ 4.85 GeV.  A sample as large as $10,000$ may be available in 
current data from runs of the CLEO detector.  

The predicted decay rates\cite{elblc} for the $\Upsilon(nS)$, $n = 1$, 3 
can be interpreted as predictions of the width for the corresponding values 
of $m_{\tilde b}$ and $m_{\tilde g}$, or as lower limits on the sparticle masses 
given known bounds on the branching fractions.  The current experimental 
uncertainties on the hadronic widths of the $\Upsilon$'s are compatible with the 
range of values of $m_{\tilde b}$ and $m_{\tilde g}$ favored in the work on the 
bottom quark production cross section in hadron reactions described above.  The 
analysis\cite{elblc} of $\Upsilon(nS)$ decays shows nevertheless that tighter 
experimental bounds on the bottom squark fraction are potentially powerful for 
the establishment of lower bounds on $m_{\tilde b}$ and $m_{\tilde g}$.   

In conventional QCD perturbation theory, the hadronic width of the $\Upsilon$ is 
calculated from the three-gluon decay subprocess, $\Upsilon \rightarrow 3g$, and 
$\Gamma_{3g} \propto \alpha^3_s$.  The SUSY 
subprocess adds a new term to the hadronic width from 
$\Upsilon \rightarrow \tilde{b} + {\tilde{b}}^*$.  If this new subprocess is 
present but ignored in the analysis of the hadronic width, the true value of 
$\alpha_s(\mu = m_b)$ will be smaller than that extracted from a standard QCD 
fit by the factor $(1 - \Gamma_{\rm{SUSY}}/\Gamma_{3g})^{\frac{1}{3}}$.
For a contribution from the $\tilde{b} {\tilde{b}}^*$ final state
that is 25\% of $\Gamma^{\Upsilon}(1S)$, the value of 
$\alpha_s$ extracted will be reduced by a factor of 0.9, at the 
lower edge of the approximately 10\% uncertainty band on the commonly 
quoted\cite{alfs} value of $\alpha_s(m_b)$.   A thorough analysis 
would require the computation of NLO contributions in SUSY-QCD to 
both the $3g$ and $\tilde{b} {\tilde{b}}^*$ amplitudes and the appropriate 
evolution of $\alpha_s(\mu)$ with inclusion of a light gluino and a light bottom 
squark.  

Direct observation of Upsilon decay into bottom squarks requires an understanding 
of the ways that bottom squarks may manifest themselves, discussed in more detail 
below.  Possible baryon-number-violating R-parity-violating decays of the bottom 
squark lead to $u+s$; $c+d$; and $c+s$ final states.  These final states of four 
light quarks should be distinguishable from conventional hadronic final states 
mediated by the three-gluon intermediate state.  For example, a greater rate 
of baryon antibaryon production is likely.  If the $\tilde{b}$ lives long 
enough, it will pick up a light quark and turn into a $B$-mesino, $\widetilde{B}$.
Charged $B$-mesino signatures in $\Upsilon$ decay include single back-to-back equal
momentum tracks in the center-of-mass; measurably lower momentum than 
lepton pairs ( $< 4$ GeV/c {\em{vs.}} $\simeq 5$ GeV/c for muons and electrons);
$1 + \cos^2 \theta$ angular distribution; and ionization, time-of-flight, 
and Cherenkov signatures 
consistent with a particle whose mass is heavier than that of a proton.  At stake 
is discovery, or new limits on the mass, of the ${\tilde b}$ as well as 
measurement of or new limits on the R-parity violation couplings of the 
${\tilde b}$.  

Pseudoscalar $\eta_b$ decay into a pair of bottom squarks is forbidden, but the 
higher-order process of $\eta_b$ decay into a pair of $B$-mesinos can proceed.
Decays of the $\chi_{b0}$ and $\chi_{b2}$ into a pair of bottom squarks are 
allowed.  Implications of a low-mass $\tilde{b}$ and low-mass $\tilde{g}$ for 
rare $B$ decay phenomena are explored by Becher {\em et al}.\cite{Kagan}

\subsection{Hadron Reactions}
Among the predictions of this SUSY scenario, the most clearcut is pair
production of like-sign charged $B$ mesons at hadron colliders, $B^+B^+$ 
and $B^-B^-$.  To verify the underlying premise, that the cross 
section exceeds expectations of conventional perturbative QCD, a new measurement 
of the absolute rate for $b$ production in run II of the Tevatron is important.   
A very precise measurement of $\bar{\chi}$ in run II is desirable.  
Since the fraction of $b$'s from gluinos changes with $p_{Tb}$, a change of 
$\bar{\chi}$ is expected when the cut on $p_{Tb}$ is changed.  The $b$ jet 
from $\tilde{g}$ decay into $b\widetilde{b}$ will 
contain the $\widetilde{b}$, implying unusual material associated with the 
$\widetilde{b}$ in some fraction of the $b\bar{b}$ data sample.  The existence 
of light $\widetilde b$'s means that they will be pair-produced in partonic 
processes, leading to a slight increase ($\sim 1$\%) 
in the hadronic dijet rate.  

The SUSY approach increases the $b$ production rate
at HERA and in $\gamma \gamma$ collisions at LEP by a small amount, not
enough perhaps if early experimental indications in these cases are
confirmed.\cite{herab,lepb}  Full NLO SUSY-QCD studies should be undertaken. 
In these two cases, the apparent discrepancy may find at least part of its 
resolution in the fact that $b \bar{b}$ production occurs very near threshold 
where fixed-order QCD calculations are not obviously reliable.  Uncertain parton 
densities of photons may play a significant role.  

\subsection{Running of $\alpha_s$}
The presence of a light gluino and a light bottom squark slow the running of the 
strong coupling strength $\alpha_s(\mu)$. Above gluino threshold, the $\beta$ 
function of (SUSY) QCD is 
\begin{equation}
\beta(\alpha_s) = \frac{\alpha_s^2}{2 \pi} \left( -11 + \frac{2}{3} n_f + 
\frac{1}{6} n_s + 2 \right) + O(\alpha_s^3).  
\end{equation}
The $\widetilde{b}$ (color triplet scalar) contributes little to the running, 
equivalent to that of 1/4'th of a new flavor,  
but the $\widetilde{g}$ (color octet fermion) is much more significant, equivalent 
to 3 new flavors of quarks.  A precise determination of $\beta(\alpha_s)$ 
appears to be the best way to confirm or exclude the possible existence of a light 
gluino.  Using a method that relies heavily on a ``renormalization group 
invariant''(RGI) technique to minimize non-perturbative and inverse-power 
contributions and scale dependence, members of the DELPHI 
collaboration\cite{DELbeta} extract the 
value $n_f^{\rm eff} = 4.75 \pm 0.44$ from an analysis of data on the thrust 
distribution $<1 - T>$ in $e^+ e^- \rightarrow {\rm hadrons}$ over the energy 
range $E_{\rm cm} = 15$ to 200 GeV.  A similar value for $n_f$ with somewhat 
larger uncertainties may be deduced from fits\cite{fits} to the $Q^2$ variation 
of $\alpha_s(Q^2)$.  These results can be compared to the pure QCD expectation 
$n_f = 5$ below $t {\bar t}$ threshold.  The small quoted uncertainty 
on $n_f^{\rm eff}$ would preclude a light gluino.  However, it remains 
crucial to understand the assumptions and constraints inherent in the use of the 
RGI approach and to verify whether the same method applied to other event 
shape-variables yields consistent results for $n_f^{\rm eff}$.  

In the standard model, a global fit to all observables provides an indirect 
measurement of $\alpha_s$ at the scale of the $Z$ boson mass $M_Z$.  The value 
$\alpha_s(M_Z) \simeq 0.1184 \pm 0.006$ describes most observables 
properly.\cite{alfs}  Extrapolation from $M_Z$ to a lower scale $\mu$ with 
inclusion of a light gluino reduces $\alpha_s(\mu)$ from its pure QCD value.  
The presence of a light gluino, with or without a light bottom squark, also 
requires reanalysis of the phenomenological determinations of 
$\alpha_s(\mu)$ at all scales to take into account SUSY processes and SUSY-QCD 
corrections to the amplitudes that describe the relevant processes.  To date, a 
systematic study of this type has not been undertaken, but, as mentioned above, 
consistency is achieved for $\Upsilon$ decays.  A lesser value of $\alpha_s(m_b)$ 
leads, under slower evolution, to the same $\alpha_s(M_Z)$.  
  

\subsection{$\widetilde b$-onia}
Bound states of bottom squark pairs could be seen as $J^P = 0^+$, 
$1^-$, $2^+$, ... mesonic resonances in $\gamma \gamma$ reactions and in 
$p\bar{p}$ formation, with masses in the 4 to 10 GeV range.  They could 
show up as narrow states in the $\mu^+ \mu^-$ invariant mass spectra at hadron 
colliders, between the $J/\Psi$ and $\Upsilon$.  At an $e^+ e^-$ collider, the 
intermediate photon requires production of a $J^{PC} = 1^{--}$ state.  Bound 
states of low mass squarks with charge $2/3$ were studied with a potential 
model.\cite{Nappi}  The small leptonic widths were found to preclude bounds 
for $m_{\widetilde{q}} > 3$ GeV.  For bottom squarks with charge $-1/3$, the 
situation is more difficult. 

\subsection{$\widetilde b$ lifetime and observability}
Strict R-parity conservation in the MSSM forbids $\widetilde{b}$ decay 
unless there is a lighter supersymmetric particle.  R-parity-violating and 
lepton-number-violating decay of the $\widetilde{b}$ into at least one lepton is 
disfavored by the CLEO data\cite{CLEO} and would imply the presence of an extra 
lepton, albeit soft, in some fraction of $b$ jets observed at hadron colliders.  
The baryon-number-violating R-parity-violating (${\not \! R_p}$) term in the MSSM 
superpotential is 
${\cal W}_{\not \! R_p} = \lambda_{ijk}^{\prime\prime}U_i^cD_j^cD_k^c$; 
$U^c_i$ and $D^c_i$ are right-handed-quark singlet chiral
superfields; and $i,j,k$ are generation indices.  The limits on individual 
${\not \! R_p}$ and baryon-number violating couplings 
$\lambda''$ are relatively weak for third-generation squarks,\cite{Allanach,BHS} 
$\lambda''_{ijk} < 0.5$ to $1$.  

The possible ${\not \! R_p}$ decay channels 
for the $\widetilde{b}$ are $123: \bar{\widetilde{b}} \rightarrow u+s$; 
$213: \bar{\widetilde{b}} \rightarrow c+d$; and 
$223: \bar{\widetilde{b}} \rightarrow c+s$.  The hadronic width is\cite{BHS} 
\begin{equation}
\Gamma(\widetilde b \rightarrow \rm{jet} + \rm{jet}) =
\frac{m_{\widetilde b}}{2\pi} \sin^2\theta_{\widetilde{b}} 
\sum_{j<k} |\lambda^{\prime\prime}_{ij3}|^2 .
\end{equation}
If $m_{\tilde b} = 3.5$ GeV, $\Gamma(\widetilde b \rightarrow i j) = 
0.08 |\lambda^{\prime\prime}_{ij3}|^2$ GeV.  Unless all 
$\lambda^{\prime\prime}_{ij3}$ are extremely small, the $\widetilde {b}$ will 
decay quickly and leave soft jets in the cone around the $b$.  Bottom-quark 
jets containing an extra charm quark are possibly disfavored by CDF, but a 
detailed simulation is needed.

If the $\widetilde{b}$ is relatively stable, the $\widetilde{b}$ could pick 
up a light $\bar{u}$ or $\bar{d}$ and become a $\widetilde{B}^-$ or 
$\widetilde{B}^0$ ``mesino" with $J = 1/2$, the superpartner of the $B$ meson.  
The mass of the mesino would fall roughly in the range $3$ to $7$ GeV for 
the interval of $\widetilde{b}$ masses favored by the analysis of the bottom 
quark cross section.  The charged mesino could fake a heavy muon if its hadronic 
cross section is small and if it survives passage through the hadron calorimeter 
and exits the muon chambers.  Extra muon-like tracks would then appear in a fraction 
of the $b \bar{b}$ event sample, but tracks that leave some activity in the hadron 
calorimeter.  The mesino has baryon number zero but acts like a heavy proton or 
antiproton -- perhaps detectable with a time-of-flight apparatus. 
A long-lived $\widetilde b$ is not excluded by conventional searches at hadron 
and lepton colliders, but an analysis\cite{baeretal} similar to that for 
$\tilde{g}$'s  should be done to verify that there are no 
additional restrictions on the allowed range of $\tilde b$ masses and lifetimes.

\section*{Acknowledgements}

I am indebted to Brian~Harris, David~E.~Kaplan, Zack~Sullivan, Tim~Tait, Carlos~
Wagner and Lou Clavelli for their collaboration.  I thank Professor Bo-Qiang Ma 
and the other members of the Local Organizing Committee for their warm hospitality 
and excellent physics discussions during the Third Circum-Pan-Pacific Symposium on 
High Energy Spin Physics, Peking University, Beijing, October 8 - 13, 2001.

\end{document}